\documentclass[conference]{IEEEtran}
\IEEEoverridecommandlockouts
\usepackage{cite}
\usepackage{amsmath,amssymb,amsfonts}
\usepackage{algorithmic}
\usepackage{graphicx}
\usepackage{textcomp}
\usepackage{xcolor}
\usepackage{tabularx}
\usepackage{pgfplots}
\usepackage{amsmath}
\def\BibTeX{{\rm B\kern-.05em{\sc i\kern-.025em b}\kern-.08em
    T\kern-.1667em\lower.7ex\hbox{E}\kern-.125emX}}
\begin{document}

\title{Speckle Noise Reduction in Ultrasound Images using Denoising Auto-encoder with Skip connection}
\author{\IEEEauthorblockN{\textsuperscript{}Suraj Bhute, Subhamoy Mandal$^{\ast}$, Debashree Guha}
\IEEEauthorblockA{\textit{School of Medical Science and Technology} \\
\textit{Indian Institute of Technology Kharagpur}\\
Kharagpur, India \\
surajrbhute123@kgpian.iitkgp.ac.in, $^{\ast}$smandal@iitkgp.ac.in, debashree\_smst@smst.iitkgp.ac.in}
}
\maketitle
\begin{abstract}
Ultrasound is a widely used medical tool for non-invasive diagnosis, but its images often contain speckle noise which
can lower their resolution and contrast-to-noise ratio. This can
make it more difficult to extract, recognize, and analyze features
in the images, as well as impair the accuracy of computer-assisted
diagnostic techniques and the ability of doctors to interpret
the images. Reducing speckle noise, therefore, is a crucial step
in the preprocessing of ultrasound images. Researchers have
proposed several speckle reduction methods, but no single
method takes all relevant factors into account. In this paper,
we compare seven such methods – Median, Gaussian, Bilateral,
Average, Weiner, Anisotropic and Denoising auto-encoder without and with skip connections - in terms of their ability to preserve features
and edges while effectively reducing noise. In an experimental
study, a convolutional noise-removing auto-encoder with skip
connection, a deep learning method, was used to improve
ultrasound images of breast cancer. This method involved
adding speckle noise at various levels. The results
of the deep learning method were compared to those of traditional
image enhancement methods, and it was found that the proposed
method was more effective. To assess the performance of these
algorithms, we use three established evaluation metrics and present
both filtered images and statistical data.

\textbf{\textit{Clinical Relevance-} Speckle noise reduction in ultrasound images is crucial for accurate diagnosis. The effectiveness of the deep learning method, auto-encoder with skip connection, in reducing speckle noise and preserving features in ultrasound images was demonstrated, leading to improved accuracy in diagnosis. This study highlights the clinical significance of this approach by enabling easier diagnosis for radiologists.}

\end{abstract}

\section{Introduction}
Ultrasound imaging is a popular choice among medical imaging techniques due to its numerous benefits. Compared to methods such as CT and X-ray, ultrasound imaging is more cost-effective, portable, and can produce real-time images without radiation.
Deep learning, a type of machine learning, allows for analysing images by using past experiences to make predictions and identify patterns. This method is particularly useful for identifying the contents of images.
Another method, auto-encoder, is an unsupervised deep learning technique that is commonly used for data compression and reducing storage space. It helps to improve system performance by removing unnecessary variables from data and can be used to visualize high-dimensional data and to remove noise from data to provide more accurate results [1,2].

 Ultrasound imaging is becoming one of the most useful techniques for breast cancer diagnosis. Actually, compared to mammography,it provides real-time imaging. In addition, it is non-invasive and does not use X-rays; is low-cost, and not generally painful. Still, one of its main disadvantages is the poor image quality, which is degraded by noise during its acquisition. Speckle spots are considered unappealing as they negatively impact the visual quality and accuracy of interpretation and diagnosis. The primary goal of denoising of image is to eliminate unwanted noise while maintaining as much important information as possible. Speckle filtering accordingly is a crucial pre-processing step for function and for better image visualization [3].
 
\section{Literature Study and Motivation}
According to [1], 2D filters for speckle removal in ultrasound images can be classified into frequency-domain filters and spatial filters. Linear spatial filters, such as wiener and median filters, can reduce noise but may cause blurring around image edges. Non-linear median-type filters aim to preserve edges, but still have limitations. Frequency-domain filters, on the other hand, effectively remove noise while preserving image edges by working on frequency information of the image. Ultrasound image denoising is an active area of research. This literature review serves as background information on the reduction of noise in ultrasound breast images using the proposed SMU (Srad Median Unsharp) algorithm. The study aims to balance the need for noise suppression and preservation of diagnostic information in medical images. The performance of the proposed algorithm is compared to other speckle noise reduction techniques and shown to have superior results [4,5]. In [5], Gondara explores the use of a convolutional autoencoder network for medical image denoising. Despite limitations of deep learning models, the proposed method proves to be efficient with a small dataset. The performance can be improved by combining heterogeneous images. Simple networks effectively reconstruct images with high levels of corruption, making noise and signal indistinguishable to the human eye [6,7]. The authors of [8] proposed a deep fully convolutional encoding-decoding framework for image restoration, including denoising and super-resolution. The network uses convolutional and de-convolutional layers for feature extraction and image detail recovery, respectively, linked by skip-layer connections. The skip connections improve training and result in better restoration performance than previous state-of-the-art methods, as demonstrated in experiments. Ye X. et al. [9] proposes a sparse denoising autoencoder method for denoising hybrid noises in images. The method is tested on natural images and evaluated using PSNR. The training process of the sparse denoising autoencoder is designed to handle single and mixed noises, making it relatively robust in practical situations. The use of autoencoder in image denoising has shown good performance and the proposed sparse denoising autoencoder model outperforms BM3D in handling hybrid noise. In [10], a patch-based image denoising method using a neural network with a convolutional autoencoder was proposed for ultra-low-dose CT images. 

In the field of ultrasound image denoising, there has been a lack of studies utilizing autoencoder with skip connection for speckle noise reduction. This study uses autoencoder with skip connection to reduce speckle noise in breast ultrasound cancer images, contributing to the understanding of its effectiveness in ultrasound image denoising.

\section{MODELING METHODS}
A denoising autoencoder with skip connections (DAE-SC) is a neural network architecture that is trained to reconstruct the original input from a corrupted version of it [7]. The DAE-SC utilizes skip connections, also known as residual connections, which bypass one or more layers in the network and directly connect the input to the output.

The mathematical equation for the DAE-SC can be represented as [9]:

$\mathbf{x}' = f(W_1 \mathbf{x} + b_1) + \mathbf{x}$

$\mathbf{\hat{x}} = g(W_2 \mathbf{x}' + b_2)$

Where,
$\mathbf{x}$ is the original input,
$\mathbf{x}'$ is the corrupted input,
$W_1$, $W_2$ are the weight matrices,
$b_1$, $b_2$ are the bias vectors,
$f(\cdot)$ and $g(\cdot)$ are non-linear activation functions,
$\mathbf{\hat{x}}$ is the reconstructed input.

The goal of the DAE-SC is to learn the weight matrices $W_1$, $W_2$ and bias vectors $b_1$, $b_2$ such that the reconstructed input $\mathbf{\hat{x}}$ is as close as possible to the original input $\mathbf{x}$. The DAE-SC is trained by minimizing the reconstruction loss between the original input $\mathbf{x}$ and the reconstructed input $\mathbf{\hat{x}}$, and the reconstruction loss is typically a mean squared error (MSE) or cross-entropy loss.

The skip connections in DAE-SC help to preserve information from the original input and prevent loss of information when passing through multiple layers, which improves the robustness and generalization of the DAE-SC.

In summary, the DAE-SC is a neural network architecture that is trained to reconstruct the original input from a corrupted version of it, it utilizes skip connections to preserve information from the original input, which improves the robustness and generalization of the network and it is trained to minimize the reconstruction loss between the original and the reconstructed input.

\section{EXPERIMENTAL STUDY}

In this study, we proposed an expanded noise-removing auto-encoder network and evaluated its performance by training it on a dataset containing various levels of noise. The experimental results were then compared with those obtained from classic image processing filters (mentioned in section B) using PSNR (Peak Signal-to-Noise Ratio), SSIM (Structural Similarity Index), and MSE (Mean Squared Error) as evaluation criteria [2].
\subsection{Data Set and Characteristics}\label{AA}

In this study, we used the Breast Ultrasound Images Dataset [11], which includes 780 images with an average resolution of 500*500 pixels in PNG format. The dataset includes both original images and their corresponding ground truth images. The images are grouped into three classes: normal, benign, and malignant. Specifically, the dataset includes 437 benign, 133 normal, and 210 malignant breast ultrasound images along with their respective ground truth images. 
\subsection{Comparison of Noise Removal Filters}

In the study, the improvement results obtained by applying the classical filters used in image processing to the test data in our dataset were compared in terms of PSNR, SSIM and MSE criteria. In the experimental study: Median, Gaussian, Bilateral, Average, Weiner, Anisotropic and Denoising auto-encoder without and with skip connections were used [1].

\subsection{Properties of the Network Used in the Experimental Study}
The parameters used in the network in the study:
\begin{itemize}

   \item{Image dimensions: 128x128}
   \item {Optimizer: Adam}
   \item {Number of revolutions (Epoch): 300}
   \item {Number of Batches: 64}
   \item {Kernel Size - 3x3}
   \item {Max Pooling - 2x2}
   \item {Learning Rate : 1e-10}
   \item {Error function: mse }
   \item {Number of training data:  546}
   \item {Number of test data: 117}
   \item {Number of test data: 117}
   \item {Monitoring : Validation Loss}
   \item {Activation Function : ReLU}
\end{itemize}

 \begin{figure}[htbp]
\centerline{\includegraphics[width=9cm]{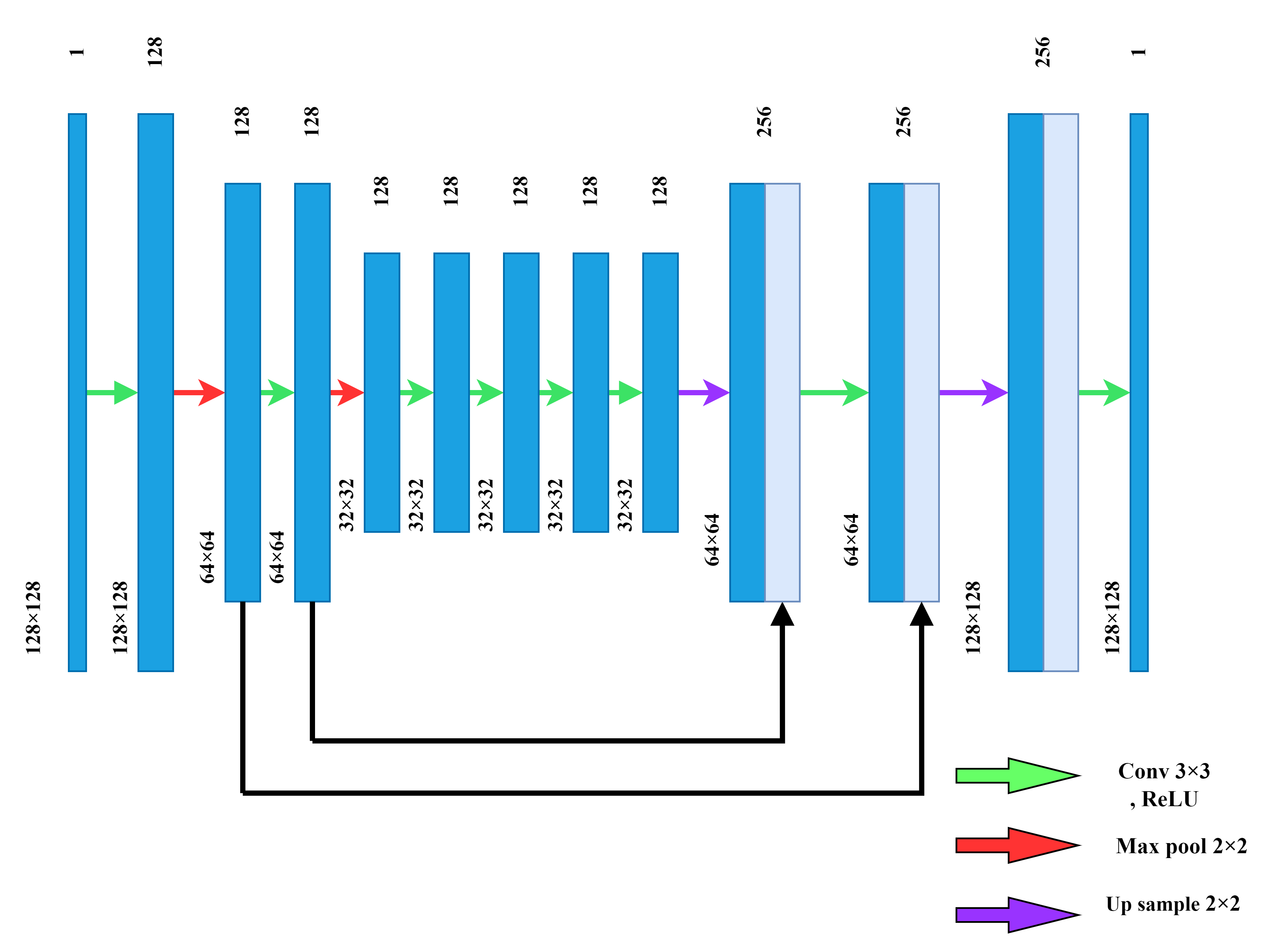}}
\caption{The Autoencoder Network with Skip Connection: The study presented an innovative autoencoder network, incorporating skip connections to remove speckle noise from ultrasound images. The network consisted of convolutional layers for feature extraction and reconstruction, and the addition of skip connections helped to preserve important information, resulting in improved performance compared to traditional filters and methods.}
\label{fig}
\end{figure}
In this study, we propose a denoising autoencoder model for image denoising. The proposed model consists of two main components: an encoder and a decoder. The encoder takes an image with noise as input and performs feature extraction through a series of convolutional and pooling layers. At each pooling layer, the image passing through the encoder is reduced in sample rate by 20\%. On the other hand, the decoder takes the encoded image and performs feature reconstruction through a series of transposed convolutional and concatenation layers. At each step, the image passing through the decoder is up-sampled by 20\% with the use of concatenation layers. The feature maps from the encoder are combined with the feature maps from the decoder, resulting in the final output of the model, which is the denoised image.
The model is trained using mean squared error loss function and the Adam optimizer. The model architecture and training details are described in this section.

\section{EXPERIMENTAL RESULTS AND DISCUSSION}
The results obtained in the study were evaluated according to PSNR, SSIM and MSE criteria. Unlike other autoencoder networks, the maximum pooling layer was used only once in the network used. The reason why the extended convolution layer is used instead of this layer is that although the maximum pooling layer has a positive contribution in terms of programming speed, it causes an increase in data loss. In addition, the use of the skip connection improved the result compared to conventional denoising autoencoder.

In the study, an autoencoder network was used for ultrasound image speckle noise reduction. It was preferred over other deep learning networks due to its efficiency in processing. The study used a single autoencoder network to train images with 5 different noise levels and obtained superior results compared to classical methods. To overcome limitations in other deep learning networks, the training dataset was created by adding different noise levels, leading to successful results in the autoencoder network.

\begin{table}[ht]
\caption{The table compares the performance of various denoising methods for different levels of noise variance (0.08, 0.1, 0.3, 0.5, and 0.7). The methods include Anisotropic, Bilateral, Weiner, Gaussian, Average, Median, Auto-encoder without skip connection and with skip connection. The performance is measured using three parameters: PSNR, SSIM, and MSE.}
\centering
\begin{tabular}{|p{2.5cm}|p{1.5cm}|p{1.5cm}|p{1.5cm}|}
\hline
Method & PSNR & SSIM & MSE \\
 \hline
 \multicolumn{4}{|c|}{Variance=0.08} \\
 \hline

Anisotropic & 11.133 & 0.081 & 0.077 \\
 \hline
Bilateral & 14.794 & 0.264 & 0.033 \\
 \hline
Weiner & 11.508 & 0.129 & 0.070 \\
 \hline
Gaussian & 14.435 & 0.232 & 0.036 \\
 \hline
Average & 14.608 & 0.248 & 0.034 \\
 \hline
Median & 14.441 & 0.241 & 0.035 \\
 \hline
Auto-encoder no skip & 20.264 & 0.900 & 0.009 \\
 \hline
\textbf{Auto-encoder skip} & \textbf{26.937} & \textbf{0.936} & \textbf{0.002} \\
 \hline
 \multicolumn{4}{|c|}{Variance=0.1} \\
 \hline
Anisotropic & 11.301 & 0.087 & 0.074 \\
 \hline
Bilateral & 14.584 & 0.252 & 0.034 \\
 \hline
Weiner & 11.242 & 0.124 & 0.075 \\
 \hline
Gaussian & 14.230 & 0.224 & 0.037 \\
 \hline
Average & 14.408 & 0.239 & 0.036 \\
 \hline
Median & 14.220 & 0.235 & 0.037 \\
 \hline
Auto-encoder no skip & 20.939 & 0.748 & 0.008 \\
 \hline
\textbf{Auto-encoder skip} & \textbf{26.555} & \textbf{0.936} & \textbf{0.002} \\
 \hline
 \multicolumn{4}{|c|}{Variance=0.3} \\
 \hline
Anisotropic & 10.828 & 0.078 & 0.082 \\
 \hline
Bilateral & 12.896 & 0.173 & 0.051 \\
 \hline
Weiner & 9.765 & 0.091 & 0.105 \\
 \hline
Gaussian & 12.629 & 0.159 & 0.054 \\
 \hline
Average & 12.772 & 0.166 & 0.052 \\
 \hline
Median & 12.575 & 0.165 & 0.055 \\
 \hline
Auto-encoder no skip & 16.729 & 0.585 & 0.021 \\
 \hline
\textbf{Auto-encoder skip} & \textbf{22.682} & \textbf{0.910} & \textbf{0.005} \\
 \hline
 \multicolumn{4}{|c|}{Variance=0.5} \\
 \hline
Anisotropic & 9.900 & 0.061 & 0.102 \\
 \hline
Bilateral & 11.961 & 0.142 & 0.063 \\
 \hline
Weiner & 8.806 & 0.068 & 0.131 \\
 \hline
Gaussian & 11.694 & 0.132 & 0.067 \\
 \hline
Average & 11.843 & 0.138 & 0.065 \\
 \hline
Median & 11.610 & 0.135 & 0.069 \\
 \hline
Auto-encoder no skip & 15.458 & 0.581 & 0.028 \\
 \hline
\textbf{Auto-encoder skip} & \textbf{21.790} & \textbf{0.919} & \textbf{0.006} \\
 \hline
 \multicolumn{4}{|c|}{Variance=0.7} \\
 \hline
Anisotropic & 9.386 & 0.058 & 0.115 \\
 \hline
Bilateral & 11.443 & 0.125 & 0.071 \\
 \hline
Weiner & 8.289 & 0.059 & 0.148 \\
 \hline
Gaussian & 11.181 & 0.112 & 0.076 \\
 \hline
Average & 11.326 & 0.119 & 0.073 \\
 \hline
Median & 11.044 & 0.116 & 0.078 \\
 \hline
Auto-encoder no skip & 15.005 & 0.619 & 0.031 \\
 \hline
\textbf{Auto-encoder skip} & \textbf{20.264} & \textbf{0.900} & \textbf{0.009} \\
\hline
\end{tabular}
\end{table}

 \begin{figure}[htbp]
\centerline{\includegraphics[width=8.5cm]{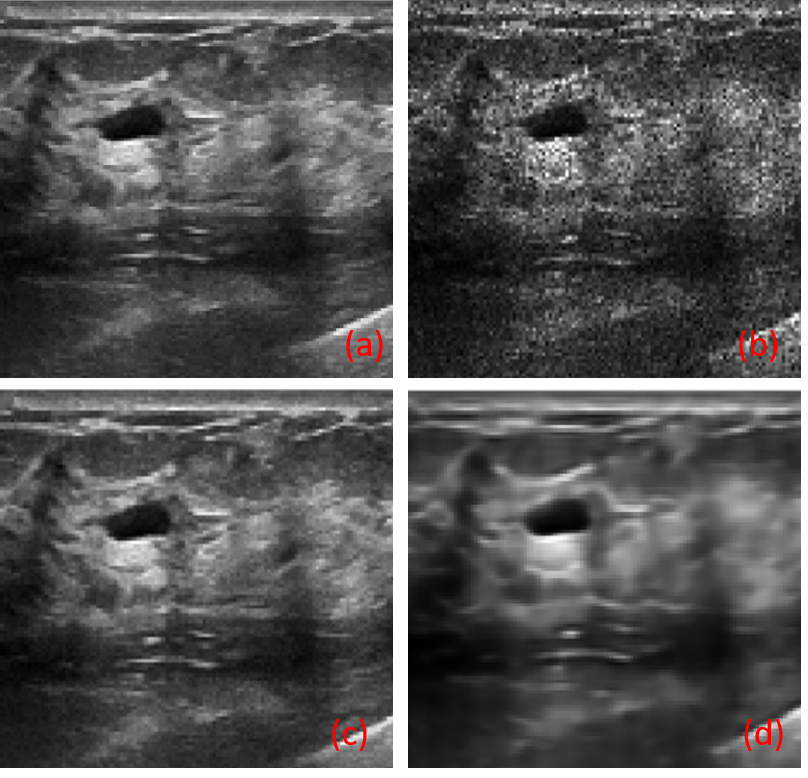}}
\centerline{\includegraphics[width=8.5cm]{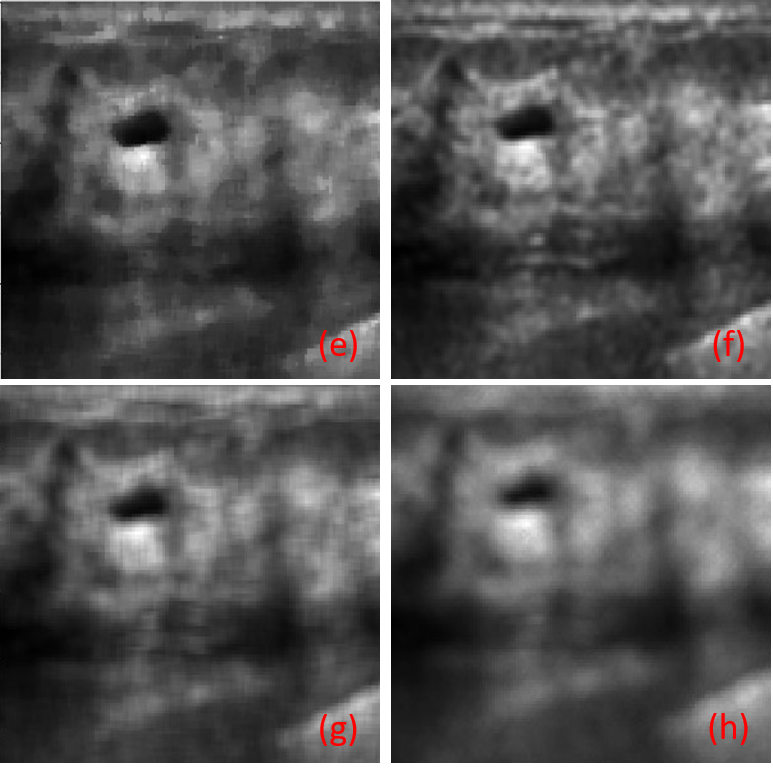}}
\centerline{\includegraphics[width=8.5cm]{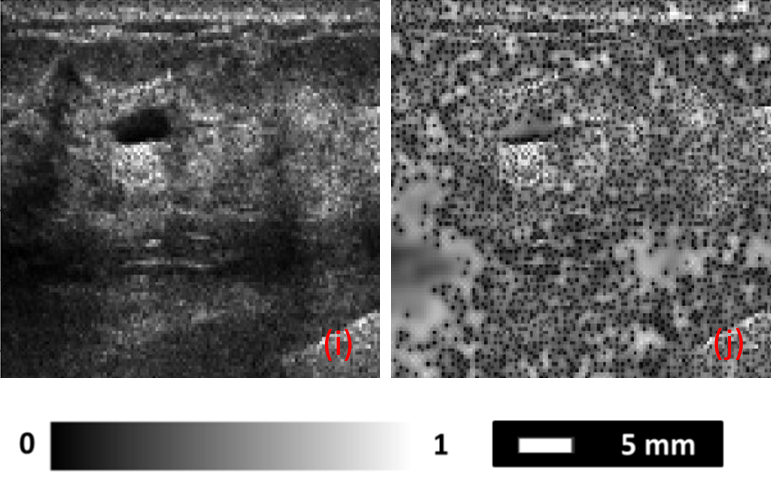}}
\caption{Comparison of different denoising techniques applied to the original image. (a) Original Image. (b) Noise image with variance 0.7. (c) Auto-encoder without skip connection. (d) Auto-encoder with skip connection, (e) Median filtering (f) gaussian filtering (g) average filtering (h) bilateral filtering (i) Weiner filter and (j) anisotropic diffusion. 
}
\label{fig}
\end{figure}

The results of the experimental study were compared in Table I. The results shown in dark color in the table correspond to the results obtained from the proposed denoising autoencoder with skip connections. As seen in the table, our proposed method yielded better results than other classical filters. The performance of the image denoising autoencoder with skip connections was evaluated using PSNR, SSIM, and MSE metrics. The network was trained with a noise variance of 0.7, resulting in a PSNR of 20.264, SSIM of 0.9, and MSE of 0.009. These results demonstrate the effectiveness of our proposed method when compared to other denoising techniques.

\begin{figure}[!htb]
\resizebox{1.0\linewidth}{!}{
\begin{tikzpicture}
\begin{axis}[
    xlabel=Variance,
    ylabel=SSIM,
    title=SSIM comparision,
    xmin=0,
    xmax=0.7,
    legend style={at={(1.05,1)},anchor=north west},
    ]

\addplot[red,mark=none] coordinates {
    (0.08,0.936) (0.1,0.936) (0.3,0.910) (0.5,0.919) (0.7,0.900)
};
\addlegendentry{Autoencoder with skip}

\addplot[green,mark=none] coordinates {
    (0.08,0.900) (0.1,0.748) (0.3,0.585) (0.5,0.581) (0.7,0.619)
};
\addlegendentry{Autoencoder without skip}

\addplot[blue,mark=none] coordinates {
(0.08,0.232) (0.1,0.224) (0.3,0.159) (0.5,0.132) (0.7,0.112)
};
\addlegendentry{Gaussian Blur}

\addplot[pink,mark=none] coordinates {
(0.08,0.248) (0.1,0.239) (0.3,0.166) (0.5,0.138) (0.7,0.119)
};
\addlegendentry{Average Blur}

\addplot[cyan,mark=none] coordinates {
(0.08,0.264) (0.1,0.252) (0.3,0.173) (0.5,0.142) (0.7,0.125)
};
\addlegendentry{Bilateral Filter}

\end{axis}
\end{tikzpicture}}
\caption{Comparison of Structural Similarity Index Measure (SSIM) using various denoising techniques: Autoencoder with skip connection, Autoencoder without skip connection, Gaussian Blur, Average Blur, and Bilateral Filter. The x-axis represents different noise levels, while the y-axis represents the SSIM values.}
\label{fig}
\end{figure}
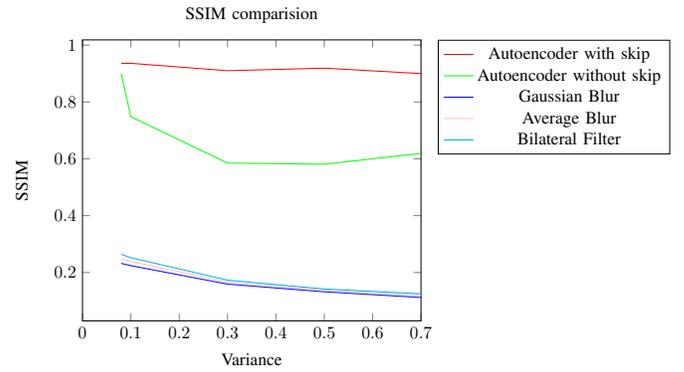

\section{CONCLUSIONS}
In deep learning networks, the network's learning capacity is expected to increase as the network becomes deeper. However, the vanishing gradient problem arises as a result of the continuous differentiation process during back-propagation, which can hinder the network's ability to learn after a certain number of cycles. This issue also affects autoencoder networks, but it can be addressed through the use of specific network architectures, such as skip connections. In this study, we have applied skip connections to overcome the vanishing gradient problem.
Our proposed denoising autoencoder with skip connections showed better results than other methods in the experimental study, Evaluated using PSNR, SSIM, and MSE metrics. These results demonstrate the effectiveness of the proposed method. In this study, the utilization of a diverse training dataset played a crucial role in enhancing the capacity of the network to learn. The dataset utilized was designed to be diverse with the aim of improving the network's performance. The results obtained demonstrate the effectiveness of this approach. However, it is anticipated that further diversification of the training dataset can lead to even better outcomes.

\vspace{12pt}
\color{red}


\begin{thebibliography}{00}
\bibitem{b1} S Pradeep, P Nirmaladevi “A Review on Speckle Noise Reduction Techniques in Ultrasound Medical images based on Spatial Domain, Transform Domain and CNN Methods” IOP Conference Series: Materials Science and Engineering
\bibitem{b} S. Paul, S. Mandal and M. S. Singh, "Noise Adaptive Beamforming for Linear Array Photoacoustic Imaging," in IEEE Transactions on Instrumentation and Measurement, vol. 70, pp. 1-11, 2021.
\bibitem{b2} Ines Njeh, Olfa Ben Sassi, Khalil Chtourou, Ahmed Ben Hamida, "Speckle Noise Reduction in Breast Ultrasound Images: SMU (SRAD Median Unsharp) Approach," 8th International Multi-Conference on Systems, Signals and Devices 2011.
\bibitem{b3} Zhang, Yifei. A Better Autoencoder for Image: Convolutional Autoencoder. In: ICONIP17-DCEC. 
\bibitem{b4} Gondara, Lovedeep. "Medical İmage Denoising Using Convolutional Denoising Autoencoders." 2016 IEEE 16th International Conference on Data Mining Workshops (ICDMW). IEEE, 2016.
\bibitem{b5} Xie, J., Xu, L., and Chen, E. Image Denoising And İnpainting With Deep Neural Networks. In Advances İn Neural İnformation Processing Systems (pp. 341-349), 2012.
\bibitem{b6} Cho, K. Boltzmann Machines And Denoising Autoencoders For İmage Denoising.
\bibitem{b7} Mao, X., Shen, C., and Yang, Y. B. Image Restoration Using Very Deep Convolutional Encoder-Decoder Networks With Symmetric Skip Connections. In Advances İn   Neural   İnformation   Processing Systems (pp. 2802-2810),2016.
\bibitem{b8}Ye, X., Wang, L., Xing, H., and Huang, L.). Denoising Hybrid Noises İn
İmage With Stacked Autoencoder. In 2015 IEEE International
Conference On Information And Automation (pp. 2720-2724). IEEE,2015, August
\bibitem{b9} Nishio, Mizuho, \textit{et.al.} "Convolutional Auto-Encoder For İmage Denoising Of Ultra-Low-Dose CT." Heliyon 3, no. 8 (2017).
\bibitem{b10}Walid Al-Dhabyani a, Mohammed Gomaa b, Hussien Khaled b, Aly Fahmy “Dataset of breast ultrasound images”.


\end{thebibliography}
\end{document}